\documentclass[aps,showpacs,prb,twocolumn,supperscriptaddress]{revtex4-1}

\usepackage{times,xspace}
\usepackage{amsbsy,amssymb,amsmath,bm}
\usepackage[dvips]{graphicx}
\usepackage{color,rotate}

\usepackage{subfigure}

\newcommand{\f}[2]{\frac{#1}{#2}}
\newcommand{\Sum}[2]{\ensuremath{\sum\limits}_{#1}^{#2}}

\newcommand{\simgeq}{\; \raisebox{-0.4ex}{\tiny$\stackrel
{{\textstyle>}}{\sim}$}\;}
\newcommand{\simleq}{\; \raisebox{-0.4ex}{\tiny$\stackrel
{{\textstyle<}}{\sim}$}\;}

\begin{document}

\title{Phase Diagram of Ferroelastic Systems in the Presence of Disorder: Analytical Model and Experimental Verification}

\author{R. Vasseur$^{1,2,3}$}
%\thanks{These authors contributed equally to this work}
\author{D. Xue$^{4}$}
%\thanks{These authors contributed equally to this work}
\author{Y. Zhou$^{4}$}
\author{W. Ettoumi$^{5}$}
\author{X. Ding$^{1,4}$}
\email[]{dingxdxjtu@gmail.com}
\author{X. Ren$^{4,6}$}
\author{T. Lookman$^{1}$}
\email[]{txl@lanl.gov}

\affiliation{${}^1$Theoretical Division and Center for Nonlinear Studies, Los Alamos National Laboratory,
Los Alamos, New Mexico 87545, USA}
\affiliation{${}^2$Institut de Physique Th\'eorique, CEA Saclay,
91191 Gif Sur Yvette, France}
\affiliation{${}^3$LPTENS, 24 rue Lhomond, 75231 Paris, France}
\affiliation{${}^4$Multi-disciplinary Materials Research Center, Frontier Institute of Science and Technology, State Key Laboratory for Mechanical Behavior of Materials, Xi'an Jiaotong University, Xi'an 710049, China} 
\affiliation{${}^5$Laboratoire de Physique des Plasmas CNRS-Ecole Polytechnique, 91128 Palaiseau
cedex, France}
\affiliation{${}^6$Ferroic Physics Group, National Institute for Materials Science, Tsukuba, 305-0047, Ibaraki, Japan}

\date{\today}

\begin{abstract}

There is little consensus on the nature of the glass state and 
its relationship to other strain states in ferroelastic materials 
which show the shape memory effect and superelasticity. We provide 
a thermodynamic interpretation of the known strain states, including
precursory tweed and strain glass, by mapping the problem onto a spin model
and analytically obtaining the 
phase diagram using real-space renormalization group methods. We further 
predict a spontaneous transition from the glass state to the 
ordered martensite phase. We verify this prediction by mapping 
out the experimental phase diagram for the ternary ferroelastic 
alloy Ti$_{50}$(Pd$_{50-x}$Cr$_x$)  and demonstrate the emergence 
of the spontaneous transition. Our work thus provides a consistent 
framework to understand the various experimental and theoretical 
studies on the glassy behavior associated with ferroelastic materials.

\end{abstract}


\maketitle

\section{Introduction}

Ferroelastic materials undergo first order transitions that are 
characterized by a lattice strain or shuffle, and transform from a 
high temperature  strain-disordered paraelastic state (austenite) to 
a low temperature  strain-ordered ferroelastic state (martensite) where long-range elastic
interactions are important. 
It is known that the thermodynamics of such phase transitions is 
strongly influenced by the presence of disorder~\cite{Ref1,Ref2,Ref3,Ref4,Ref5,Ref6,Ref7}. In 
particular, statistical compositional fluctuations play a fundamental 
role in bringing about a precursory strain state known as  tweed. This 
is a crosshatched pattern observed well above the martensitic-transformation 
start temperature~\cite{Ref8,Ref9,Ref10}. Recent experiments 
on ferroelastic alloys have shown that by introducing disorder via doping
point defects or compositional variations beyond  
a critical value, an abnormal glass-like state which is a frozen 
state of local strain order, can be generated below a  
transition temperature~\cite{Ref11,Ref12,Ref13,Ref14,Ref15}. This glass phase was initially observed in Ni-rich
Ti$_{50-x}$Ni$_{50+x}$ where the austenitic $B2$ parent structure appeared
to persist to 0 K above a compositional threshold $x > 1.3$, below which it transformed to a
 martensitic $B19'$ phase~\cite{Ref11}. This so 
called ``strain glass'' is of interest not only from a theoretical 
point of view but it has been shown to display superelasticity 
and shape memory effects, which are typically seen in austenite 
and martensitic states~\cite{Ref15}. 

Experiments and theory have so far provided little understanding of 
the nature of the glassy behavior in ferroelastic materials and its 
relationship to other strain states. For example, the tweed has been 
the subject of a number of theoretical studies and has variously been 
interpreted as a ``glass'' phase~\cite{Ref9,Ref10}. Although recent experiments 
distinguish strain glass from tweed~\cite{Ref12}, the claims are largely based 
on diagnostics that monitor certain static and dynamic aspects of 
the materials; these include the broken ergodicity of static 
properties ({\it e.g.}, the strain measured by zero field and field 
cooling (FC-ZFC)), or frequency dispersion of dynamic properties~\cite{Ref14}. 
However, non glassy phases, such as polytwins in martensite or 
nanostructured ferromagnetics ({\it e.g.} FeReCr), also show similar 
non-ergodic static behavior and frequency dispersion~\cite{Ref16,Ref17}. 
In addition, although numerical simulations based on continuum 
Landau descriptions in the presence of disorder~\cite{Ref13} can reproduce 
some form of the experimental results~\cite{Ref11,Ref12,Ref13,Ref14,Ref15}, such solutions are 
also based on similar empirical diagnostics and tend not to be 
predictive. Therefore, there is a need for a predictive approach 
using analytical techniques that would allow the various ordered 
and disordered ferroelastic states to be distinguished and which 
can be experimentally verified.

In this paper, we provide a thermodynamic interpretation of the known 
strain states in a ferroelastic system, including austenite, martensite, tweed and strain glass. After mapping the problem onto
a spin model, we use real space renormalization  
group (RG) methods to analytically calculate thermodynamic phases in 
terms of RG attractive fixed points. The RG approach progressively 
integrates out microscopic degrees of freedom so that the attractive 
basins characterize the physics at large scales. The values of the 
interaction strength and strength of disorder uniquely characterize 
the different phases, including the frustrated glassy state, thereby 
allowing the phase diagram in terms of temperature and disorder to 
be determined. We thus theoretically determine the phase diagram of a model ferroelastic, 
and the phase diagram predicts a spontaneous transition from strain 
glass to martensite if the strength of disorder in the system has 
intermediate values. This latter aspect was not recognized previously, although the spontaneous
transition has been recently speculated~\cite{Ref18}. 
 We verify these predictions by mapping out the experimental 
diagram for a ternary ferroelastic Ti$_{50}$(Pd$_{50-x}$Cr$_x$) alloy (where $x$  
is the atomic concentration of Cr) and examining its consequences. Moreover, our calculations show
 that the tweed phase can be interpreted as a thermodynamic equilibrium phase.
Our model thus provides a unified, consistent framework to understand 
the various experimental and theoretical studies on the glassy behavior 
associated with ferroelastic materials. Moreover, we find that
the long range elastic interaction or the precise form of the disorder are not  crucial to  describing the phase diagram.
These ideas are well recognized for spin glasses but are not as well
known  within the strain glass community.

The plan of our paper is as follows. In the first section, we review
how effective spin models can be deduced from Ginzburg-Landau (GL) functionals, and we discuss the various means to include
quenched disorder in the model.
The second section is devoted to the calculation of the
analytical phase diagram from our model in the presence of disorder. We explain how a real-space RG procedure can be implemented and we present
several approaches and approximations to solve the resulting equations. 
The results are compared to Monte-Carlo simulations and we provide a discussion 
of the relevance of simplified numerical algorithms -- be it a one-spin flip Monte-Carlo or 
the steepest-descent method for continuum GL models-- for these types of  problems.
The role of the long-range interactions is also discussed in relation to 
previous work.  Section~\ref{Section-Exp} contains our experimental results on Ti$_{50}$(Pd$_{50-x}$Cr$_x$) alloys. 
We measure the transformation behavior of the alloys by means of dynamical mechanical analysis (DMA) 
and the electric resistivity. In-situ synchrotron X-ray and transmission electron microscopy (TEM) were 
employed to detect the predicted spontaneous phase transformation. Our experiments lead to  a phase diagram that is very  
similar to the one we establish analytically. Finally, we discuss our results in the broader context of ferroelastic
transitions in general and the possible generalizations of our theoretical work
to more complex transitions.

\section{Pseudo-spin model and quenched impurities}
 
Our approach is to use a model that captures the salient physics and 
microstructure associated with ferroelastic transformations. We will use a
pseudo-spin model derived from a continuum formulation based on a 
Landau potential and strain compatibility forces~\cite{Ref19,Ref20}. The 
addition of quenched disorder then permits studies of the model 
using the tools of statistical mechanics. We review in this section
the crucial steps in the derivation of such spin models starting
from standard Ginzburg-Landau (GL) free energy functionals.
%We then introduce the real-space RG formalism on the simple example
%of the pure (without quenched disorder) spin model. 
We then discuss how the effect of impurities in the original GL theories
can be taken into account in spin models.
 
\subsection{Spin model} 
\label{subsecPureSpinmodel}

The idea of using a spin model for ferroelastic transitions was introduced 
in Ref. ~\cite{Ref19}. This idea was later generalized to many other transitions
and shown to capture the salient physics of continuum GL models~\cite{Ref20}.
For simplicity, we will study here 2D square to rectangle (SR) transformation 
driven by the deviatoric strain $e_2 = \frac{1}{\sqrt{2}} (\epsilon_{11} - \epsilon_{22}) $, 
where $\epsilon_{\mu \nu}$ are components of the strain tensor 
$\epsilon_{\mu \nu}=\frac{1}{2}(\frac{\partial u_{\mu}}{\partial r_{\nu}} + 
\frac{\partial u_{\nu}}{\partial r_{\mu}})$ defined in terms of displacements, $u$.
The generalization to more complicated -- and more realistic -- transitions
will be discussed in Sec.~\ref{SectionGeneralization}.
The Landau free energy of the system is $F_{\rm GL} \left[e_2 \right] = E_0 \int d^2 r \left[f_L + f_G + f_{LR} \right] $, 
where the local free energy for the first-order transition is $f_L=(\tau -1) e_2^2 + e_2^2 (e_2^2-1)^2$, 
the gradient term $f_G = \xi^2 \left| \nabla e_2 \right|^2$, accounts for the cost 
of creating interfaces between different variants and 
$f_{LR}= \frac{A_1}{2}\int d^2 r' e_2(\vec{r}) U(\vec{r}-\vec{r}') e_2(\vec{r}') $  is the elastic long-range force. 
The scaled temperature $\tau = \frac{T-T_c}{T_{eq}-T_c}$ is expressed 
in terms of the transition temperature  $T_{eq}$ and $T_c$ is the temperature of the austenite stability limit. 
The long-range term comes from the so-called compatibility equation $\vec{\nabla}\times\left(\vec{\nabla}\times\mathbf{\epsilon}\right)^T = 0$ on the
strain tensor $\mathbf{\epsilon}$, that ensures that the displacement field is single valued.
This leads to the anisotropic Kernel $U(\vec{r}-\vec{r}')$, which reads in Fourier space 
\begin{equation}
\hat{U}(\vec{k})=\f{\left({k_x}^2-{k_y}^2\right)^2}{k^4+8A_1{k_x}^2{k_y}^2/A_3} \,\nu(\vec{k}).
\end{equation}
The factor $\nu(\vec{k})=1-\delta_{k,0}$ ensures that this long-range term vanishes for a uniform strain field.
In all what follows, the ratio $A_1/A_3=\frac{1}{2}$ will be kept constant.
Back into real space we roughly have $U \simeq \frac{\cos 4 \theta}{r^2}$.

The strain $e_2$ is the order parameter for this transition. The Landau term $f_L=(\tau -1) e_2^2 + e_2^2 (e_2^2-1)^2$
ensures that the system undergoes a first-order phase transition at $\tau=1$. When $\tau<4/3$, this term has
three minima, which correspond to phases called austenite ($e_2=0$) and martensite ($e_2=\pm\varepsilon(\tau)$), where
$\varepsilon(\tau)=\left[\f{2}{3}\left(1+\sqrt{1-\f{3\tau}{4}}\right)\right]^{1/2}$. For $1<\tau<4/3$, the austenite is a metastable state.
To obtain non-uniform textures, one minimizes the GL functional $F_{\rm GL} \left[e_2 \right]$ as usual.
Note that at this stage the GL continuum model is in itself a mean-field description of the microscopic problem,
and as such it does not contain any information about fluctuations around the mean-field solution given by the minimization
of $F_{\rm GL} \left[e_2 \right]$. An alternative point of view is to consider a formal partition function $\mathcal{Z}=\int 
\mathcal{D}[e_2] \mathrm{exp}\left(-\beta F[e_2]\right)$, where the strain configurations are summed over with a weight
given by the usual Boltzmann weight. The Laudau (or Mean-field) approximation is then nothing but a saddle point calculation
of this functional integral. One can simplify the calculation of $\mathcal{Z}$ by retaining only the strain minima given 
by the Landau part of $F$, transforming then $e_2$ into a discrete variable that we rewrite as $e_2(\vec{r}) = \varepsilon(\tau) S(\vec{r})$
with the pseudospin $S(\vec{r})=0,\pm1$. This is the main point of the pseudo-spin approximation, instead of 
keeping only one configuration which minimizes the full GL functional $F_{\rm GL} \left[e_2 \right]$, we
retain in the partition function all the configurations that minimize the local Landau term $f_L=(\tau -1) e_2^2 + e_2^2 (e_2^2-1)^2$.
Within this approximation, the partition function reads $\mathcal{Z} \simeq \sum_{\lbrace S\rbrace} \mathrm{exp}\left(-\beta H \right)$,
where the Hamiltonian $H$ is obtained from the continuum theory as $ H = F_{\rm GL} \left[e_2(\vec{r})  \rightarrow \varepsilon(\tau) S(\vec{r})  \right]$.
After a proper discretization~\cite{Ref19,Ref20}, it reads
\begin{equation}
\beta H = - J(\tau) \sum_{<i,j>} S_i S_j + \Delta(\tau) \sum_i S_i^2 + \frac{\beta A_1}{2} \sum_{ij} S_i U_{ij} S_j,
\label{eq_Hpure}
\end{equation}
where $\Delta(\tau) = D_0(\tau)/2 \left( g_L(\tau) + 4 \xi^2 \right)$ and $J(\tau)= D_0(\tau) \xi^2$, 
with $D_0=2\beta E_0\varepsilon(\tau)^2$, $g_L=(\tau-1)+(\varepsilon^2(\tau)-1)^2$.
The states $S_i=\pm 1$ correspond to the two rectangular variants while $S_i=0$ represents 
to the square austenite. Note that this spin model only makes sense for $\tau \leq 4/3$,
as for $\tau > 4/3 $, only the state $S=0$ is allowed.
We have therefore mapped our continuum GL theory onto a lattice spin$-1$
model with Hamiltonian~\eqref{eq_Hpure}. This Hamiltonian with $A_1=0$ is known in the spin literature 
as the Blume-Capel model~\cite{Ref21} and it reproduces all the well-known features of the transition~\cite{Ref20},
in particular, it has a first-order transition around $\tau \simeq 1$.

There are several points that are worth mentioning at this stage.
It is usual to deduce GL theories from lattice models, not the other way around. The reason for this is that
GL theories are easier to deal with and contain interesting coarse-grained, mean-field information about the initial 
microscopic model. In our case, we chose to
start from a simplified GL theory  to map onto a spin-$1$ statistical model that may seem harder to handle.
On the other hand, spin models will turn out to be very convenient when  introducing quenched disorder 
 as connections to usual spin glass models can then be made~\cite{Ref8,Ref22}.
It is also important to realize that when progressing from mean-field GL theory 
to a classical spin model we have actually put in additional information about thermal fluctuations, as
the former {\it a priori} does not contain any information about fluctuations around the mean-field solution.
The Hamiltonian~\eqref{eq_Hpure} is an {\it effective} model, with temperature-dependent coefficients and 
 is {\it not} an accurate microscopic description of the phenomenon that we are studying.
The additional information contained in the effective Hamiltonian~\eqref{eq_Hpure} is somehow arbitrary
and other choices of reasonable Hamiltonian would be possible. For example, one could as well have introduced two different
temperatures, the physical temperature $T$ that would appear in $\tau$, and another more artificial temperature $T_e$ that would appear
in the Boltzmann factor $\exp (-H/T_e)$. As was suggested in Ref.~\cite{Ref8}, one can then also let $T_e = 0$
so that one would need to minimize $H$ in order to obtain statistical properties. 
We choose to introduce somewhat artificially thermal fluctuations by taking $T=T_e$, a point of view 
completely analogous to what is usually done with the well known $\phi^4$ theory. This is a crucial point when
dealing with such effective spin models.

\subsection{Spin Model with Quenched Disorder}

Now that we have seen how to model ferroelastics using spin models, we consider the 
influence of impurities  in the model. Our aim is to understand how quenched 
impurities affect the phase diagram of ferroelastic materials. This is usually done 
phenomenologically in Laudau theory~\cite{Ref13} by introducing a random field or by introducing 
randomness in the transition temperature $T_c$
with non-zero spatial correlations~\cite{Ref23,Ref24}. In this paper, we follow the idea that
one need not worry about the precise microscopic form of the disorder induced by impurities
as one expects the effect of quench disorder in the interactions of the model to yield somewhat 
universal features. Recall that our spin model is an effective model, so the results we are
after are generic features of the phase diagram, such as its topology, rather than an accurate, quantitative
description to be compared directly with experiments. Note that this train of thought is very similar 
to what was done in spin glass theory for usual ferromagnets (see {\it e.g.}~\cite{Ref25,Ref26, Ref27}). 
The point is then to start from the pure (disorder-free) Hamiltonian~\eqref{eq_Hpure} and to 
take the nearest-neighbor couplings to be quench independent random variables $J_{ij}$,
drawn from the distribution $\mathcal{P}(J_{ij})$, with mean $J(\tau)$ and with variance $\sigma_J$.
As we will see later, the precise form of this distribution is irrelevant to the global 
topology of the phase diagram. This is a strong argument in favor of the ``universality'' 
discussed previously. We believe that this way of introducing disorder tends to be more satisfying than
the very specific, sometimes fine-tuned, methods usually used in the literature. 
The parameter $\sigma_J$ can be thought of as a measure of the quenched 
disorder in the system, in particular, for $\sigma_J=0$ we recover a pure system.
For future reference, we give the explicit form of our disordered Hamiltonian
\begin{equation}
\beta H = - \sum_{<i,j>} J_{ij}(\tau) S_i S_j + \Delta(\tau) \sum_i S_i^2 + \frac{\beta A_1}{2} \sum_{ij} S_i U_{ij} S_j.
\label{eq_Hdisorder}
\end{equation}
where the couplings $J_{ij}$ are quenched variables drawn from the distribution $\mathcal{P}(J_{ij})$.
In this paper we will use two different types of distributions
\begin{subequations}
\begin{align}
 \mathcal{P}(J_{ij}) &= \frac{1}{2} \delta(J_{ij}-J_1)+\frac{1}{2} \delta(J_{ij}-J_2),  \label{eqn:Bimodal}\\
 \mathcal{P}(J_{ij}) &=  \frac{1}{\sqrt{2 \pi} \sigma_J} \exp \left(- \frac{(J_{ij}-J(\tau))^2}{2 \sigma^2_J} \right), \label{eqn:Gaussian} 
\end{align}
\end{subequations}
with $J_1=J(\tau)+\sigma_J$ and $J_2=J(\tau)-\sigma_J$.
In the following, the distribution~\eqref{eqn:Bimodal} will be referred to as ``Bimodal'' whereas we will denote~\eqref{eqn:Gaussian}
as Gaussian. These equations define the model we will attempt to solve in the next section.

\section{Phase Diagram}

%After discussing the previous attempts to solve the disordered version of the spin model~\eqref{eq_Hdisorder},
We now consider how real-space Renormalization-Group (RG) can be used
to compute the full phase diagram of the the spin model~\eqref{eq_Hdisorder}.
We also present preliminary Monte Carlo results and discuss their relationship
to the phase diagram of the model.

\subsection{Real-Space Renormalization Group for the pure model}
\label{subsecRGpure}

Our goal in this section is to obtain an analytic phase diagram of the model defined by 
eq.~\eqref{eq_Hpure} in the limit $A_1=0$. From now on we will only take the limit
of vanishing long-range interactions $A_1=0$ as we do not believe they are important
to understand the phase diagram of ferroelastic materials. We will return to this
later. There are several ways to deal with a Hamiltonian such as~\eqref{eq_Hpure},
an obvious straightforward method would be to perform a mean-field approximation. 
This yields (see {\it e.g.} Ref.~\cite{Ref20} for a description in the context 
of martensitic transitions) a first order phase transition at $\tau \simeq 1$
between a martensite phase with $m=\langle S\rangle \neq 0$ and an high-temperature 
austenite phase characterized by $m=0$. Here we choose a different, usually more reliable method:
the real-space Renormalization Group (RG). Its application to our pure spin-$1$ model
goes back to Ref.~\cite{Ref28}, and is of course well known. 
We review here some of the essential elements before proceeding to the case of disorder.
 
The idea is to approximate the square lattice by a hierarchical lattice, obtained by iterating the construction scheme 
\begin{equation} \label{eqn:cstr}
\includegraphics[width=8.0cm]{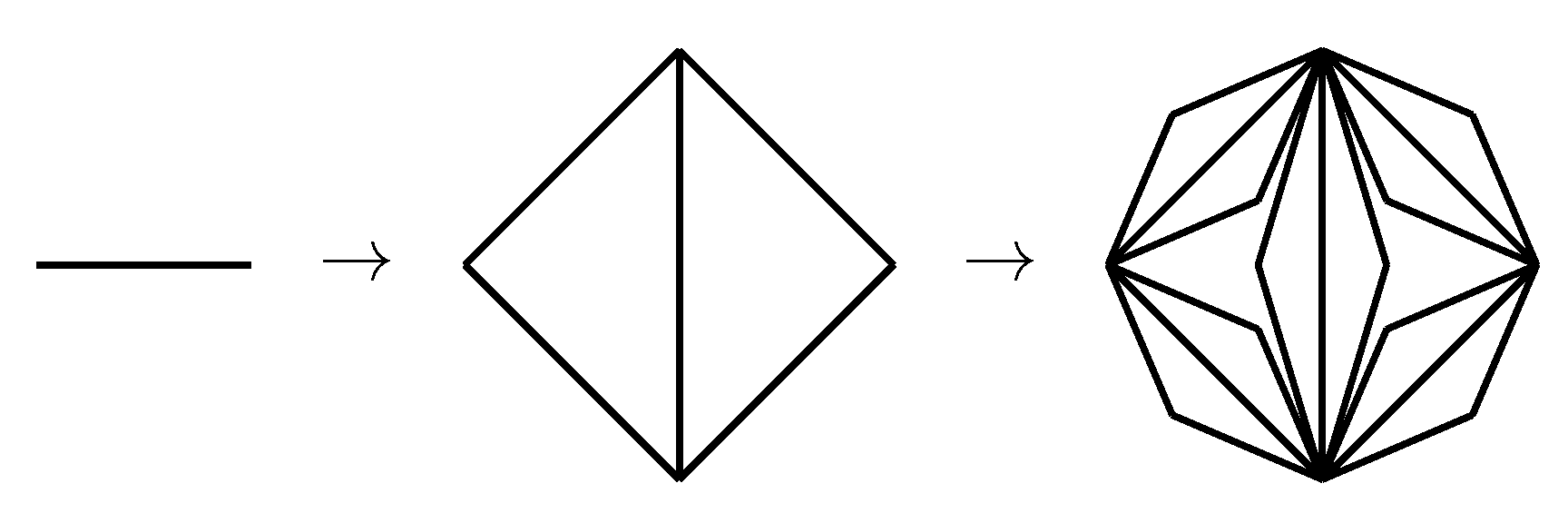}
\end{equation}
Several other choices of hierarchical lattices  to approximate the square lattice are possible,
however the one we choose is known to be a very good approximation in the case of spin-$1$ models (see {\it e.g.} Ref.~\cite{Ref29}).
We define our spin model on such a lattice by the Hamiltonian $\beta H = -J \sum_{<i,j>} S_i S_j + \frac{\Delta}{4} \sum_{<i,j>} (S_i^2+S_j^2) $.
Note that we expressed the crystal field term $\Delta$ as a sum over the links instead of a sum over sites
in order to take into account the fact that the coordination number of the sites in the lattice~\eqref{eqn:cstr}
is not constant. It is well-known that such precautions are needed when dealing with 
the renormalization of local field terms such as $-h \sum S_i$ or specifically here $\Delta \sum S_i^2$.
This corresponds to assigning weights to the sites in the crystal-field interaction term ($\Delta$) 
according to their coordination numbers which are, unlike the square lattice, not uniform. 
This inhomogeneity is necessary so that the square lattice is correctly 
approximated using a hierarchical lattice.

On such a lattice, the partition function can be computed exactly by summing progressively over the spins.
The renormalization procedure consists in decimating spins according to the  scheme
\begin{equation} \label{eq_RGstep}
\includegraphics[width=8.0cm]{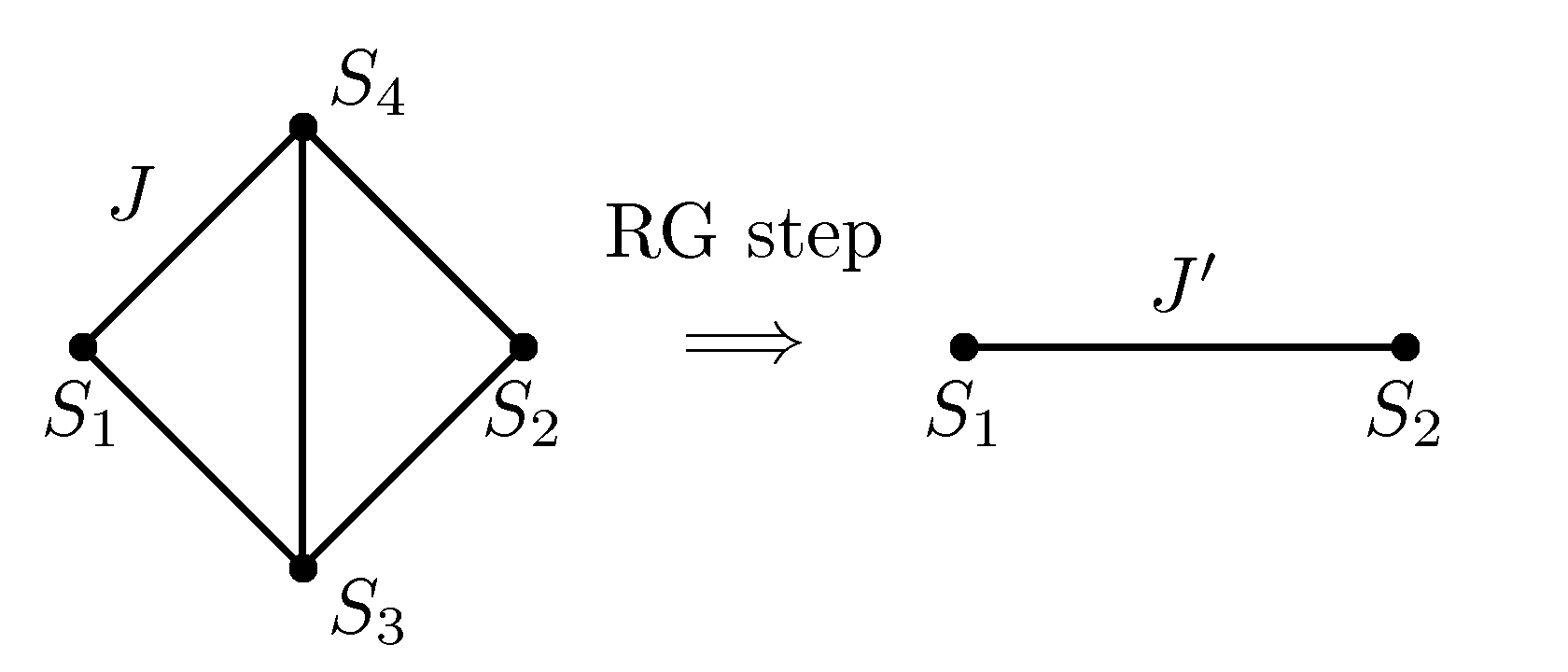}
\end{equation}
so that the renormalization process corresponds to inverting the arrows in eq.\eqref{eqn:cstr}.
More formally, if one denotes by $H_{1234}$ the Hamiltonian of the left-hand side of eq~\eqref{eq_RGstep} and by $H^{\prime}_{12}$ its right-hand side, one has
\begin{equation}
\mathrm{exp}\left[-\beta H^\prime_{12}\right] = \mathrm{Tr}\: \mathrm{exp}\left[-\beta H_{1234}\right],
\label{eqn:RGstepW}
\end{equation}
where $\mathrm{Tr}$ amounts for a partial summation performed on spins $S_3$ and $S_4$.
This equation then gives the relationship between $H_{1234}$ and the renormalized Hamiltonian $H^\prime_{12}$, 
so that one can draw the evolution of the coupling parameters by repeating this decimation procedure. 
It is not hard to see that this transformation is exact if and only if one introduces an additional coupling $- K \sum_{<i,j>} S_i^2 S_j^2$
that is generated upon renormalization. Although this biquadratic interaction is absent in our original model,
it is generated by the RG procedure, and must be taken into account to follow the exact RG flow.
In our case, the elementary Hamiltonian $\mathcal{H}_{1234}$ reads
$\beta H_{1234} = - \Sum{\left\langle i,j\right\rangle}{} \left[ J S_i S_j + K {S_i}^2 {S_j}^2\right] 
+ \frac{1}{4} \Delta \Sum{\left\langle i,j \right\rangle}{} \left[ {S_i}^2 + {S_j}^2\right]$,
where the sum are performed over the links ($\left\langle13\right\rangle$,$\left\langle14\right\rangle$,$\left\langle42\right\rangle$,$\left\langle32\right\rangle$,$\left\langle34\right\rangle$) in eq.~\eqref{eq_RGstep}.
It is worth emphasizing again that the additional coupling term $-K \sum_{\left\langle i,j\right\rangle} {S_i}^2 {S_j}^2$ must be 
introduced in the Hamiltonian, otherwise there is no analytical 
solution to be found for eq.~\eqref{eqn:RGstepW}. The renormalized Hamiltonian $\mathcal{H}^\prime_{12}$ then reads
$\beta H^\prime_{12} = - J^\prime S_1 S_2 - K^\prime {S_1}^2 {S_2}^2 + \frac{\Delta^\prime}{4} \left({S_1}^2 +  {S_2}^2 \right) + C^\prime$,
where $C^\prime$ becomes a simple multiplying constant $\exp(-C^\prime)$ that contributes to the renormalization of the 
free energy, irrelevant for our purposes.

It is straightforward to solve eq.~\eqref{eqn:RGstepW} for all $S_1$,$S_2$.
This yields a non-linear map $(J',\Delta',K') = \mathcal{R} \left[J,\Delta,K \right]$ that gives
the exact renormalized couplings of the effective Hamiltonian $H'$ in terms of the initial ones. Iterating this map, one can
obtain the effective Hamiltonian describing the physics of the system at large scale, 
where the microscopic degrees of freedom of the system have been summed over. Basins of attraction 
of this map correspond to thermodynamical phases, characterized by attractive fixed points.
For example, the martensite phase (ferromagnetic phase in the spin language) has $J^*=+\infty$ and $\Delta^*=-\infty$,
so at large scales the effective Hamiltonian describing the system forces the spin to be in the $S=\pm 1$ states,
with a strong correlations between nearest neighbors. Meanwhile, austenite can be described by the attractive fixed
point $J^*=0$ and $\Delta^*=\infty$ so it corresponds to a disordered phase that favors $S=0$.
To obtain the phase diagram of our spin model~\eqref{eq_Hpure} for $A_1=0$, we simply iterate 
the map $\mathcal{R}$ starting from the initial point $(J,\Delta,K)=(J(\tau),\Delta(\tau),0)$.
We find that the system flows to the martensite fixed point for $\tau \simleq 0.96$, and to 
the austenite fixed point for $\tau \simgeq 0.96$. Moreover, 
the study of the largest eigenvalue of the linearized
Renormalization-Group matrix around the transition at $\tau \simeq 0.96$ can be used to
prove that the transition is of first order~\cite{Ref28,Ref29}, as expected.

\subsection{Real-space RG with Disorder}
\label{subsecRGdisorderedcase}

The idea of using spin glass models to describe disordered ferroelastic materials
originates from Refs.~\cite{Ref8,Ref22}.
The replica/Mean-Field approach of this model~\cite{Ref22,Ref30} yields a strain glass phase
but also several unwanted features that are contrary to both experiments and
Monte-Carlo simulations. The absence of a tweed precursor is an example.
We use here a different approach relying on real-space RG
which we believe is simpler and usually more reliable. Although our starting
point is the same Hamiltonian~\eqref{eq_Hdisorder}, we will see in
this paper that  RG provides interesting new results that can 
be tested experimentally.

The implementation of the RG procedure for disordered systems is very similar
to the pure case in Sec.~\ref{subsecRGpure}.
The real-space RG in disordered systems still relies on a decimation procedure to progressively 
sum over the spin variables, integrating out the microscopic degrees 
of freedom (see {\it e.g.} Refs.~\cite{Ref28,Ref29,Ref31}). 
After each summation step, the effective Hamiltonian 
describing the remaining degrees of freedom is assumed to be the 
same as the original one but with renormalized coupling coefficients.
This turns out to be exact for the hierarchical lattice~\eqref{eqn:cstr}. 
The only difference is  that for disordered systems the RG 
allows us to follow the probability distribution of the couplings upon the 
renormalization procedure instead of the couplings themselves. 
In order to do so, the main point is to obtain the RG equations in the case of a
completely inhomogeneous system. Even though the initial Hamiltonian~\eqref{eq_Hdisorder} (with $A_1=0$)
has pure $K=0$ and $\Delta$ interactions, the RG flow will lead to randomness in all couplings,
so one has to consider the more general Hamiltonian
$
\beta H = - \Sum{\left\langle i,j\right\rangle}{} \left[ J_{ij} S_i S_j + K_{ij} {S_i}^2 {S_j}^2 
 - \frac{\Delta_{ij}}{4} (S_i^2+S_j^2) - \frac{\Delta^\dagger_{ij}}{4} (S_i^2-S_j^2) \right]. $
The initial condition for the RG flow is the Hamiltonian~\eqref{eq_Hdisorder} with $A_1=0$,
so that before renormalization, $K_{ij}=0$, $\Delta_{ij}=\Delta({\tau})$ and $\Delta^{\dagger}_{ij}=0$.
Remark that we once again considered the crystal field term as living on the edges of the lattice,
and not on the sites. This should be reminiscent of our treatment of the pure case in Sec.~\ref{subsecRGpure}.
The solution of eq.~\eqref{eqn:RGstepW}
in this case yields a map $\mathcal{R}$ that gives the renormalized couplings 
$(J_{12}^\prime,K_{12}^\prime,\Delta^\prime_{12},\Delta^{\dagger \prime}_{12})$
in terms of the original ones $(J_{14},J_{13}, \dots, K_{14}, K_{13}, \dots, \Delta_{14},\Delta_{13}, \dots,
\Delta^\dagger_{14},\Delta^\dagger_{13}, \dots)$. 
%It is important to emphasize again that as one iterates this map, any randomness in the initial couplings $J_{ij}$ will lead
%to random renormalized couplings $\Delta^\prime$ or  $K^\prime$.
%All couplings thus have to be considered as random variables.
One then analyzes the evolution of the joint probability distribution
$\mathcal{P}(J_{ij},K_{ij},\Delta_{ij},\Delta_{ij}^\dagger)$
upon renormalization, starting from the initial distributions
$\mathcal{P}(\Delta_{ij}) = \delta(\Delta_{ij}-\Delta(\tau))$, $\mathcal{P}(\Delta^\dagger_{ij}) = \delta(\Delta^\dagger_{ij})$,
$\mathcal{P}(K_{ij}) = \delta(K_{ij})$ and $\mathcal{P}(J_{ij})$ given by either~\eqref{eqn:Bimodal} or~\eqref{eqn:Gaussian}. 
Let us now be slightly more explicit. 
Let ${\bf K_{ij}} = (J_{ij},K_{ij},\Delta_{ij},\Delta^\dagger_{ij})$ denote the four couplings living on the link $(ij)$.
The RG maps ${\bf K_{ij}}^\prime=\mathcal{R}[\lbrace {\bf K_{ij}} \rbrace]$ then gives 
the 4 renormalized couplings ${\bf K_{ij}}^\prime = (J_{12}^\prime,K_{12}^\prime,\Delta^\prime_{12},\Delta^{\dagger \prime}_{12})$ as a 
function of the 20 initial ones $\lbrace {\bf K_{ij}} \rbrace = ({\bf K_{13}},{\bf K_{14}},{\bf K_{34}},{\bf K_{23}},{\bf K_{42}})$,
where we used the labeling of eq.~\eqref{eq_RGstep}. 
The RG recursion relation for the joint distribution 
$\mathcal{P}({\bf K_{ij}})=\mathcal{P}(J_{ij},K_{ij},\Delta_{ij},\Delta^\dagger_{ij})$
then reads
\begin{equation}
\mathcal{P}^\prime({\bf K_{ij}}^\prime) = \int \left[ \prod_{ij} {\rm d} {\bf K_{ij}} \mathcal{P}({\bf K_{ij}}) \right] \delta \left({\bf K_{ij}}^\prime-\mathcal{R}[\lbrace {\bf K_{ij}} \rbrace] \right), 
\label{RGeqDistributions}
\end{equation}
where the product is over the five links of eq.~\eqref{eq_RGstep}, so there are 20 integrations
to be performed at each RG step.
We emphasize again that even though one starts from pure $K$ and $\Delta$ couplings,
randomness in these parameters will be generated by the RG procedure. Similarly, 
even if one starts from a bimodal Gaussian distribution for $\mathcal{P}(J_{ij})$,
the resulting distribution obtained after several renormalization steps could
 in principle be much more complicated.

\subsection{Basins of attraction and thermodynamical phases}

\begin{table}
\begin{center}
\begin{tabular}{|c|c|c|}
  \hline
  Phase & OP characterization & RG fixed point  \\
  \hline
  Austenite & $m=q=0$, $p$ small & $\Delta^*=+\infty$, $J^*=0$, $\sigma^*_J=0$ \\
  Martensite & $m \neq 0$, $p \neq 0$, $q \neq 0$ & $\Delta^*=-\infty$, $J^*=\infty$, $\frac{\sigma^*_J}{J^*}=0$ \\
  Tweed & $m=q=0$, $p$ large & $\Delta^*=-\infty$, $J^*=0$, $\sigma^*_J=0$ \\
  Strain glass & $m = 0$, $p \neq 0$, $q \neq 0$ & $\Delta^*=-\infty$, $J^*=0$, $\sigma^*_J=\infty$ \\
  \hline
\end{tabular}
\end{center}
\caption{Characterization of the different thermodynamic phases in terms of the 
calculated Renormalization Group (RG) fixed points. The austenite and tweed phases are 
clearly separated because one fixed point has value $\Delta^{*}=+\infty$ (so the state $S=0$ is favored) whereas the other has $\Delta^{*}=-\infty$. 
We also show the corresponding OP values  $m=\overline{\left< S_i \right>}$, $p=\overline{\left< S_i^2 \right>}$, and $q=\overline{\left< S_i \right>^2}$  for the phases as characterized in mean field and replica theory. }
\label{tabOP}
\end{table}

Before solving explicitly eq.~\eqref{RGeqDistributions}, let us discuss how the phases 
are characterized within this formalism.
The RG distribution $\mathcal{P}(J_{ij},K_{ij},\Delta_{ij},\Delta^\dagger_{ij})$ will typically flow to 
attractive fixed points that will characterize thermodynamical phases. In principle,
one would need an infinite number of parameters to characterize RG-invariant distributions
$\mathcal{P}^*(J_{ij},K_{ij},\Delta_{ij},\Delta^\dagger_{ij})$. However, for our purposes it turns out that the thermodynamic 
phases can obtained as RG basins of attraction characterized 
only by the values of $\Delta^*$ and ($J^*,\sigma_J^*$) at the fixed point. 
Here $\Delta^*$ and $J^*$ are the mean values of $J$ and $\Delta$ at the fixed point,
while $\sigma_J^*$ is the standard deviation of $J$. 
%Within the projection approximation, these couplings are the natural parameters
%that appear in the imposed distributions.

The martensite (ferroelastic) phase corresponds to the usual ordered ferromagnetic 
phase $(J^*=\infty, \Delta^*=-\infty,\sigma^*_J/J*=0)$. In terms of order parameters 
(OPs) used in mean field theory and replica calculations (see {\it e.g.}~\cite{Ref22}), this phase is characterized by a 
non-zero magnetization $m=\overline{\left< S_i \right>} \neq 0$, where the bar represents average 
over the disorder and the brackets average with respect to Boltzmann weights. 
We find two paraelastic disordered phases with $J^*=0$, $\sigma_J^*=0$, and $\Delta^* = \pm \infty$. 
The case $\Delta^* =\infty$ corresponds to the austenite case as it favors $S=0$ whereas 
the case $\Delta^* =-\infty$ is  interpreted as a disordered phase of martensite clusters 
which we identify as the tweed phase.
The OP that allows to distinguish between both phases is the ``martensite volume fraction'' 
$p=\overline{\left< S_i^2 \right>}$. 
%Note that these  two ``paraelastic'' phases
%already exist in the pure system~\cite{Ref18} ($\sigma_J=0$).
We also remark that the tweed precursor we find is ergodic and non-glassy, consistent with recent experiments~\cite{Ref32}.
This is to be compared with the hypothesis of Refs~\cite{Ref9,Ref10} that interpreted tweed as a glassy phase. Our model results
are that tweed is a thermodynamic phase, rather than a metastable precursor. 
Note also that this tweed phase is not captured by a mean-field/replica-symmetric analysis of our 2D model~\cite{Ref22}.
The last phase that we encounter corresponds to a spin/strain glass with infinite 
randomness ($\sigma^*_J = \infty$) and ($J^*=0$, $\Delta^* = \infty $). 
The effective Hamiltonian describing the system at large scales has features in which variants  $S=\pm 1$ are 
favored (because $\Delta^* = \infty $), 
and the values $J^*=0$ and $\sigma^*_J = \infty$ imply random infinite 
couplings on each bond, thus denoting frustration. This phase is also characterized by the Edwards-Anderson 
order parameter $q=\overline{\left< S_i \right>^2}$ which in the replica language corresponds to the overlap 
between two replicas $q=\left< S_i^{1} S_i^{2} \right>$ of the system~\cite{Ref25}. 
The characterization in terms of RG fixed points or OPs of these four thermodynamic phases is gathered in Tab.~\ref{tabOP}.

\begin{figure}[tbph] 
    \includegraphics[width=8.0cm]{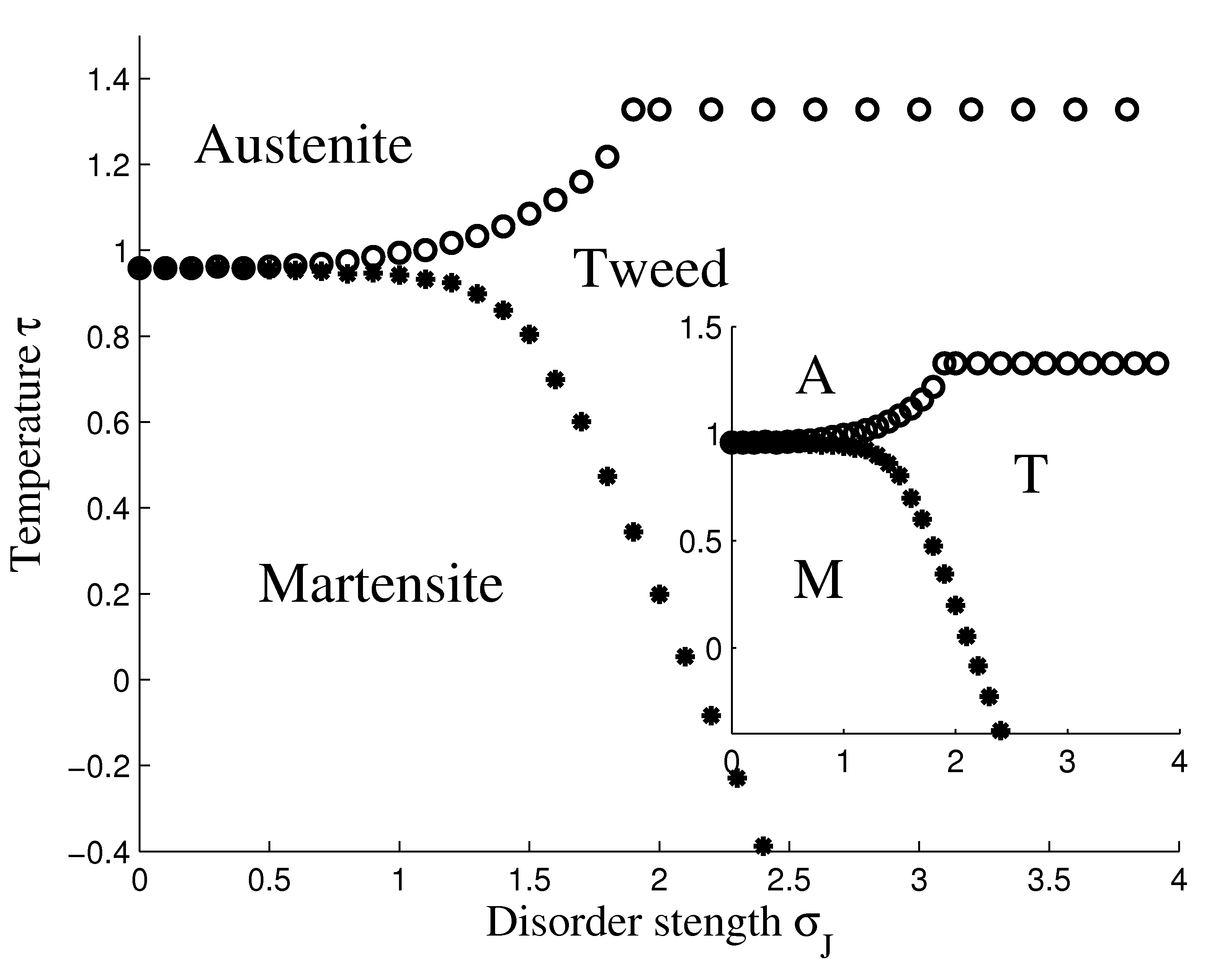}
   
  \caption{Analytical phase diagrams from the exact numerical resolution of the RG equations on the hierarchical lattice. 
  Main figure: the initial distribution of the disorder was taken to be bimodal . Inset: initial Gaussian distribution.
  Notice the absence of spin/strain glass phase (see text).}
  \label{figFullRG}
\end{figure}

\subsection{Analytical phase diagram}

\subsubsection{Numerical Resolution}

We now discuss how one can solve~\eqref{RGeqDistributions}.
An obvious method would be to sample numerically the joint distribution 
$\mathcal{P}(J_{ij},K_{ij},\Delta_{ij},\Delta^\dagger_{ij})$. This allows one to solve explicitly 
eq~\eqref{RGeqDistributions} numerically. 
We used pools of 160000 values to sample the distribution $\mathcal{P}(J_{ij},K_{ij},\Delta_{ij},\Delta^\dagger_{ij})$. 
Results are shown in Fig.~\ref{figFullRG}
for both bimodal~\eqref{eqn:Bimodal} and Gaussian~\eqref{eqn:Gaussian} initial distributions.
Parameters appropriate for our example are $E_0 = 3$, $\xi = 0.5$, $ T_{eq} = 1$, and $T_c = 0.9$. 
Note that strictly speaking, our spin model is not defined for $\tau>4/3$, as the spin approximation
only yields one state $S=0$ in that case. However, we can still think of this region as being in the austenite
phase, as this is the only phase allowed by the spin approximation.
One of the most striking features of these phase diagrams is the appearance of an intermediate tweed phase
between the austenite and martensite phases as the disorder is turned on. This is consistent
with what is known experimentally (see experimental section). 
Meanwhile, the phase diagram has the same topology
for both types of disorder. This is a strong result in favor of the ``universality'' 
discussed in Sec.~\eqref{subsecPureSpinmodel}.

One notes the absence of the spin/strain glass phase, which is to be expected. 
Strictly speaking, the model~\eqref{eq_Hdisorder} in two dimensions should not
have a spin-glass phase at finite temperature~\cite{Ref31}.
To be more precise, the Hamiltonian~\eqref{eq_Hdisorder} is believed
to have a spin glass lower-critical dimension $d_c$ lying somewhere between 2 and 3.
This means that for $d<d_c$ (low dimension), and in particular for $d=2$, the spin glass 
phase should be destroyed by thermal fluctuations. Meanwhile, it was
recently proved~\cite{Ref31} that the very same model (forgetting about the temperature
dependence of the coefficients) has a spin glass phase at finite temperature in 3D. (Note that the SR transition embedded in 3D -- or our spin model in 3D -- corresponds to a slightly constrained tetragonal to orthorhombic transition).
Of course, a spin glass phase can also be found within a mean-field/replica
solution of the model~\cite{Ref22}, as mean-field effectively corresponds to $d=\infty$.
We expect this spin glass phase to reappear on higher-dimensional 
hierarchical lattices~\cite{Ref31}. Although one might think that the absence of 
spin glass phase in 2D could be of crucial importance experimentally,
this is actually irrelevant to us, for the following reasons:
\begin{itemize}
\item Even though a spin glass phase should not exist in 2D, we still
expect kinetic features to make the system look ``glassy''. This will be discussed
in more detail in the Monte Carlo section below. The point is that the distinction in experiments and numerics 
between a genuine spin glass phase and a glassy kinematic behavior is actually
a very subtle issue. 
\item As our spin model was derived from a mean-field Landau energy without thermal fluctuations, 
we are interested in generic (mean-field 
like) features of our model. This is a consequence of the discussion at the end of 
Sec.~\eqref{subsecPureSpinmodel}. The spin glass 
lower critical dimension is not a meaningful quantity in our case as our model
was derived from a mean-field Landau energy in the first place.
Therefore, the fact that the somewhat artificial fluctuations -- introduced in going from the continuum
GL theory to our spin model-- may or may not destroy the spin glass phase is 
not relevant , as these fluctuations do not exist in the original Landau model.
Our model is thus only meaningful in a mean-field-like context.
\end{itemize}  
Therefore, one need not worry about this issue of lower critical dimension here,
it would be meaningful only if our spin model were a precise microscopic description of ferroelastics.
  
  \begin{figure*}[tbph] 
\begin{center}
    \includegraphics[width=16.0cm]{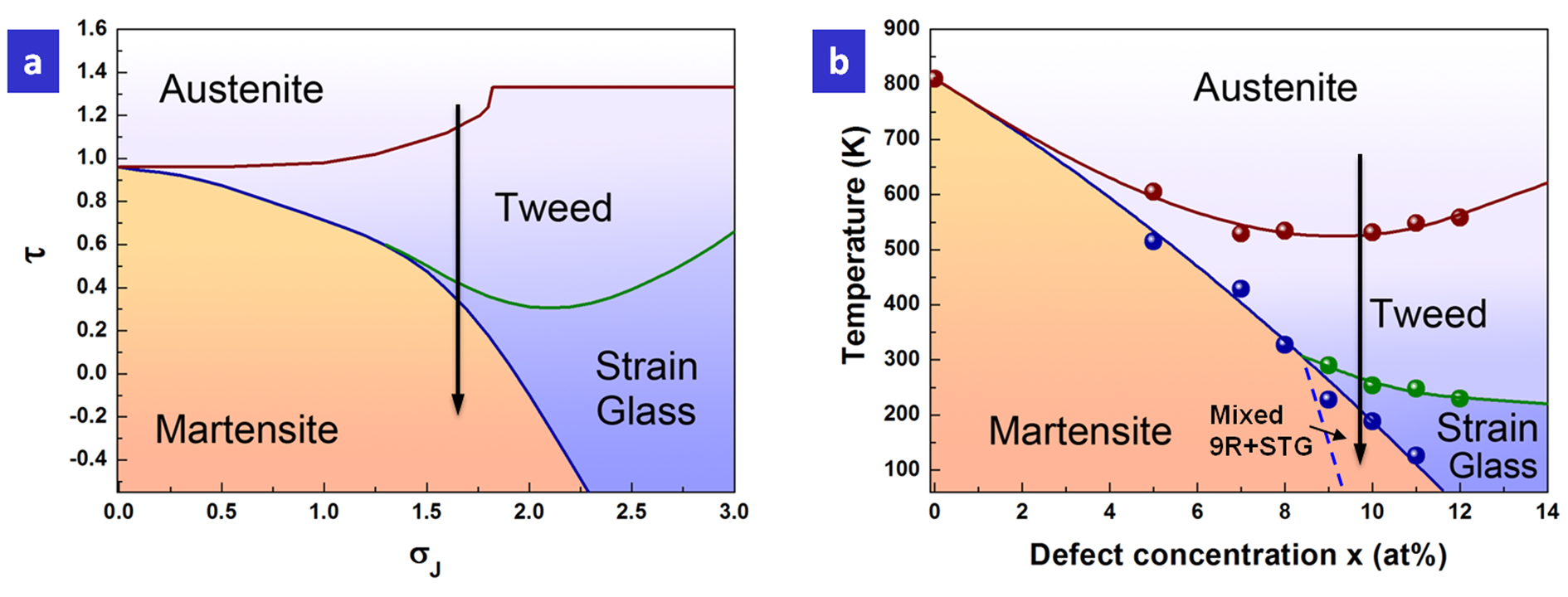}
   
  \caption{ Comparison between theoretical and experimental phase diagram. (a) 
  Phase diagram in the temperature-disorder ($\tau$,$\sigma_J$) plane for our spin model, obtained within the RG projection approximation.  $\tau$ is 
  the normalized temperature and $\sigma_J$ characterizes the amount of quenched disorder 
  in the system. (b) Experimental phase diagram of the ternary ferroelastic Ti$_{50}$(Pd$_{50-x}$Cr$_x$).}
  \label{fig1}
  \end{center}
\end{figure*}

\subsubsection{Projection Approximation}

Although the direct numerical resolution of eq.~\eqref{RGeqDistributions} 
is the most straightforward way to proceed, it is also useful to have an approximate
way of solving this system, thereby allowing one to perform the calculations analytically.
A possibility to avoid this rather cumbersome numerical procedure
is to make a further approximation in the case of Bimodal disorder~\eqref{eqn:Bimodal}.
One can remark that rather than follow the full evolution of these distributions, one could enforce 
the renormalized distributions to be the same as the initial ones but with 
renormalized parameters. That is, we enforce the distribution of the $J_{ij}$
couplings to remain Bimodal upon renormalization, keeping $K$ and $\Delta$ constant.
Although this approximation may appear too drastic, it is quite 
common in spin glass related problems and we
will show in the following that it captures all the important
features of the phase diagram of the model~\eqref{eq_Hdisorder}.
Note that such an approximation typically yields 
results characteristic of higher-dimensional systems (see {\it e.g.} Ref.~\cite{Ref29} 
and references therein), and thus should alter the spin glass lower critical dimension. 
As discussed previously, this is not important for our purposes.
In the following, we will refer to this approximation
as the ``projection approximation'', as it indeed consists in projecting the renormalized
distributions onto the initial ones.

Fig.~\ref{fig1}(a) shows the phase diagram in the plane $(\tau, \sigma_J)$ obtained by iterating the RG map 
in the projection approximation case. In the absence of disorder ($\sigma_J = 0$), we find a first-order phase 
transition between the austenite and martensite phase with $\tau \simeq 1$, as expected. As one increases the disorder, 
an intermediate tweed phase arises before it transforms into low temperature phase (either 
martensite or strain glass).
For large enough disorder and low temperature, we find a spin 
glass phase that we interpret as a strain glass. Interestingly, when the disorder of the 
system is in the intermediate regime ($1.3< \sigma_J <2.3$ in our model), we find there 
should exist a spontaneous phase transition from strain glass to martensite, this is a 
prediction that was not obtained by previous numerical 
simulations~\cite{Ref13,Ref19,Ref23,Ref24,Ref33} based on Landau theory.
The existence of a spin glass phase in this calculation
is related to our projection approximation, which is known to produce results 
characteristic of higher-dimensional systems (it is therefore
legitimate to think of this approximation as a kind of mean field). 
Note also that except for the spin-glass phase,
the exact numerical resolutions of Fig.~\ref{figFullRG} and the projection approximation results
of Fig.~\ref{fig1}(a) are very much alike.

\subsection{Monte Carlo simulations and influence of Long-range interactions}

\begin{figure*}[tbph] 
\begin{center}
    \includegraphics[width=18.0cm]{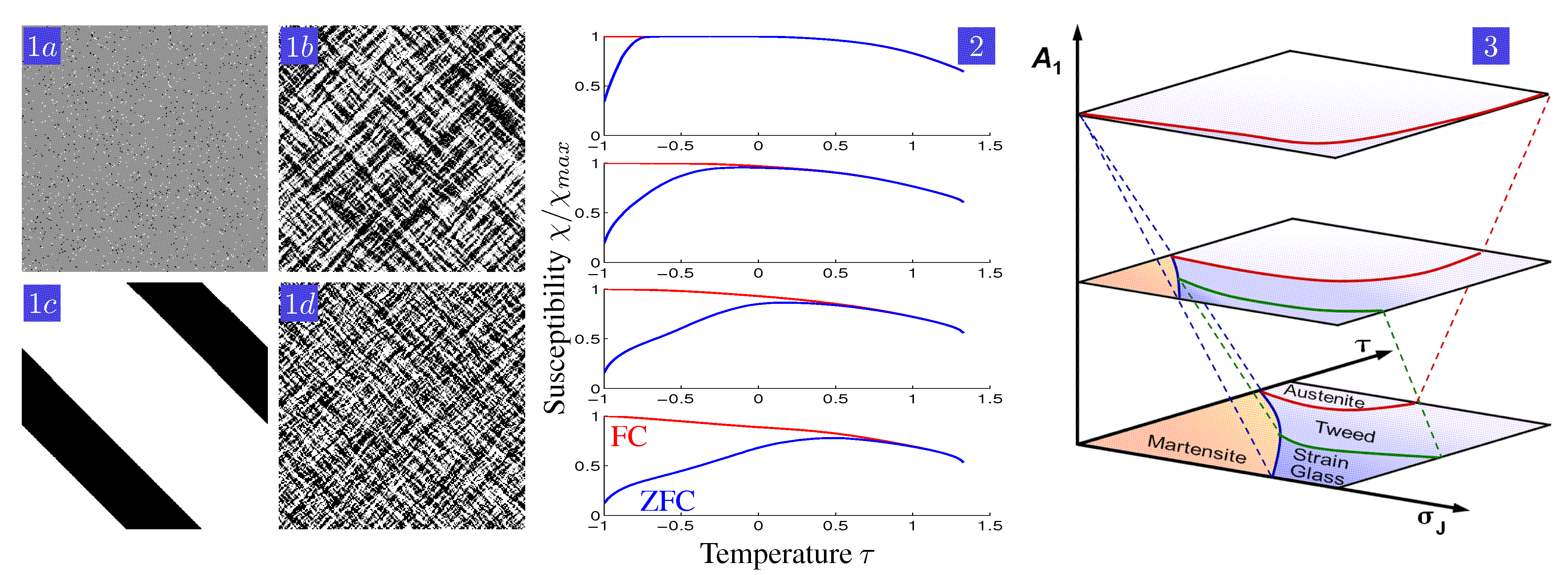}

\caption{Monte Carlo resuts, parameters are given in the text.(1) Typical microstructures obtained for $A_1=4$ on a $256 \times 256$ lattice in the different phases of the phase diagram. (1a) austenite, (1b) tweed, (1c) martensite, (1d) ``strain glass''. (2) Example of FC/ZFC curves with $A_1=0$ for disorder $\sigma_J$=1.5, 2, 2.25, 2.5 from top to bottom. The curves represent the (normalized) susceptibility $\chi = m/h$
against the temperature $\tau$. (3) Qualitative phase diagram showing the influence of the long range interaction  and disorder on the various phase transitions. Four different phases are shown: austenite, martensite, tweed and strain glass. } 
\label{FigMC}
\end{center}
\end{figure*}

We also present some preliminary Monte Carlo (MC) simulations that tend to confirm our RG results.
We first note that one has to be very careful when doing MC simulations on a disordered system,
in particular, one has to use algorithms that properly sample the configurations of the system, such
as Replica MC or parallel tempering~\cite{Ref34}. Unfortunately, these algorithms lose
their full efficiency because the coefficients in our Hamiltonian are temperature-dependent. 
We thus use a simplified simulated annealing algorithm, which we expect to yield reasonable
results at least for small disorder. The algorithm is the
analog of the steepest descent method used in the literature to minimize disordered
GL functionals. We  emphasize here that such minimization methods are often inadequate 
when dealing with spin-glass-like systems as the free-energy typically possesses many metastable minima.
We first describe microstructure obtained in the presence of long-range interactions. 
%Although we obviously do not think this is a satisfactory criterion, this is an important 
%point because such microstructures are usually by themselves
%considered as characteristic signatures of the phases of the system.  
Fig~\ref{FigMC} (a) shows typical microstructures obtained in different regions of the phase diagram on a $256 \times 256$ lattice,
with the strength of the long-range term of $A_1=4$. These textures are fully consistent with what is observed in continuum GL theories and in
experiments. In particular, we find the usual cross-hatched pattern for the tweed phase. Nevertheless, 
 our RG approach provides a clear meaning to the tweed phase, even in the absence of long-range interactions.
We also show in Fig.~\ref{FigMC} (b) Field-cooling/Zero-field-cooling results (FC/ZFC) that are usually used
in both numerical studies and experiments to test for breaking of ergodicity and glassiness. 
These curves were obtained by averaging over $~10^3$ disordered configurations on $64 \times 64$ lattices,
other lattice sizes were tested without much difference.
The curves shown in Fig.~\ref{FigMC} (b)
may be interpreted as a signature of history dependence or ergodicity breaking. However, as argued  previously, we do not expect the 2D version
of the Hamiltonian~\eqref{eq_Hdisorder} to have a spin-glass phase. The word ``spin-glass'' should be understood here in the technical sense of the term, this does not prevent the system from showing kinematic ``glassy behaviors'' in the sense sometimes used by experimentalists. 
Therefore, our ZFC/FC  results should be interpreted as a pure kinematic
effect, possibly related to the slow convergence of our MC algorithm. Similar problems should occur when minimizing GL disordered functionals 
thanks to naive steepest descent algorithms. It is worth emphasizing that probing
spin-glass phases in usual spin models is a very subtle question to address numerically, and this cannot be answered using simple
ZFC/FC experiments.

Finally, we suggest that the main features of our phase diagram persist even in the presence of long-range interactions (see related discussions in Refs~\cite{Ref23,Ref24}).
The influence of the elastic long-range interaction on the austenite/tweed and tweed/strain glass transition temperatures is shown schematically in Fig.~\ref{FigMC}(3). For no disorder ($\sigma_J=0$), the austenite/martensite transition temperature decreases linearly with $A_1$, as included phenomenologically within Landau theory. All the transition temperatures decrease with $A_1$, in particular the glass transition is shifted to lower temperatures because the long range interactions compete with the randomness~\cite{Ref22}. In the asymptotic limit $A_1 \rightarrow \infty$, the disorder becomes irrelevant and only the austenite phase remains; we therefore conjecture that the phase diagram is shifted to lower temperatures with increasing $A_1$. This result could have implications in the study of colossal magneto-resistance (CMR) materials where the interplay of disorder and long-range strain mediated interactions has a bearing on phase separation of coexisting insulating and conducting phases~\cite{Ref35}.

\subsection{Extension to three dimensions}

\label{secGene3D}

We now put our work focused on the 2D SR transition in a more general context and  explain how 
one can extend the results to more realistic situations. 
Although our RG analysis is in 2D and is for a square to rectangle transition (SR), we do not expect the salient features of the phase diagram to change for other transitions in 2D or 3D.  
As an example, let us discuss how one can extend our approach to the 3D Cubic to Tetragonal transition (CT). A spin model for this transition was already proposed in Ref.~\cite{Ref20}. 
The CT transition is described in terms of a two-dimensional order parameter (OP) given by the adjunction of both deviatoric and shear distortions $(e_2=\left(\epsilon_{xx}-\epsilon_{yy}\right)/\sqrt{2},e_3=\left(\epsilon_{xx}+\epsilon_{yy}-2\epsilon_{zz}\right)/\sqrt{6})$.
The free energy functional can be written as a sum of 3 terms, 
$F=\int \mathrm{d}^3 \vec{r} \left[f_L(e_2,e_3)+f_G(\vec{\nabla}e_2,\vec{\nabla}e_3)+ \frac{A_1}{2}f_{LR}(e_2,e_3)\right]$.
The Ginzburg term reads $f_G = \xi^2 (\left| \nabla e_2\right|^2+\left| \nabla e_3\right|^2) $, whereas $f_{LR}(e_2,e_3)$ is a long-range
part that shall not be important here (its explicit expression can be found in Ref.~\cite{Ref20}). 
The Landau part is slightly more complicated $f_L(e_2,e_3)=\tau({e_2}^2+{e_3}^2)-2({e_3}^3-3e_3 {e_2}^2)+({e_2}^2+{e_3}^2)^2$.
This free energy can be minimized with respect to $e_2$ and $e_3$, leading to four minima for $\tau < \frac{9}{8}$. 
Let alone the $e_2=e_3=0$ minimum, the three other ones are given in the complex plane by $\varepsilon(\tau){\omega_3}^k$, 
with ${\omega_3}^3=1$, and where similarly to the 2D case, one introduces
\begin{equation}
\varepsilon(\tau)=\f{3}{4}\left(1+\sqrt{1-\f{8\tau}{9}}\right).
\end{equation}
One can observe that $\tau=4/3$ now corresponds to the upper spinodal.
Retaining the Landau minima in the free energy $f_L$, we define a pseudospin $\vec{S}$ such that
$\vec{e}=(e_2,e_3)^{T} \rightarrow \varepsilon(\tau) \vec{S} $, with
\begin{equation}
\vec{S}\in\left\{\left(
\begin{array}{ll}
0\\
0
\end{array}
\right),
\left(
\begin{array}{ll}
1\\
0
\end{array}
\right),
\left(
\begin{array}{ll}
-1/2\\
\pm \sqrt{3}/2
\end{array}
\right)
\right\}
\end{equation}
The Landau part of the free energy hence reduces to
$
f_L(\tau)=\varepsilon^2(\tau)g_L(\tau)\vec{S}^2(\vec{r}),
$
where
$
g_L(\tau)=\tau-1+\left(\varepsilon^2(\tau)-1\right)^2.
$
This leads to the pseudospin model 
\begin{equation}
\label{eq3DHpure}
\beta H = - J(\tau) \sum_{\langle i,j \rangle} \vec{S}_i.\vec{S}_j + \Delta(\tau) \sum_{i} \vec{S}_i^2 + \frac{\beta A_1}{2} \sum_{ij} U_{ij} \vec{S}_i.\vec{S}_j  ,
\end{equation} 
with $\Delta(\tau)=\frac{D_0(\tau)}{2}(g_L(\tau)+6 \xi^2)$, $J(\tau)=D_0(\tau) \xi^2$, and $D_0(\tau)$ defined as before.
This model is a three-dimensional clock model with long-range interactions. The two-dimensional spin $\vec{S}$ can take 3
values on the unit circle in addition of the value $\vec{S}=\vec{0}$.

It is straightforward to extend the renormalization group method to this class of models.
This is done by considering hierarchical lattices with fractal dimension close to 3, for which
the RG decimation step becomes exact if one introduces an additional coupling $-K \sum_{\langle i,j \rangle} (\vec{S}_i.\vec{S}_j)^2$.
We solved exactly the pure model defined by eq.~\eqref{eq3DHpure} for $A_1=0$, on a three-dimensional hierarchical lattice to find
 two phases separated by a first order phase transition around $\tau \simeq 1$, as expected.
The high-temperature austenite phase is characterized by $\langle \vec{S}\rangle = \vec{0}$, whereas
the martensite phase  shows one of the three variants lying on the circle $|\vec{S}|=1$.
Quench disorder can be easily introduced through random $J_{ij}$ couplings and the RG iteration
can be  generalized to the disordered case.
The calculated phase diagram is very similar to that obtained for SR in 2D in Fig.~\ref{fig1}(a), although the numerical resolution of the RG equations is much more difficult in this case. We find two different ``paraelastic''
phases, the usual austenite characterized by the RG fixed point ($J^*=0$, $\Delta^* =\infty$), and a tweed
fixed point with ($J^*=0$, $\Delta^* =-\infty$) which corresponds to a disordered phase of martensitic variants.
We also expect this kind of model to show a spin/strain glass phase, at least in high dimensions.
These results are consistent with our extensive study of the SR 2D model. It is thus very tempting to conjecture
that the behavior of even more complicated clock models, describing more evolved transitions, 
should have a phase diagram very similar to that of the SR model.

\section{Experimental phase diagram}

\label{Section-Exp}

\begin{figure*}[tbph] 
\begin{center}
    \includegraphics[width=18.0cm]{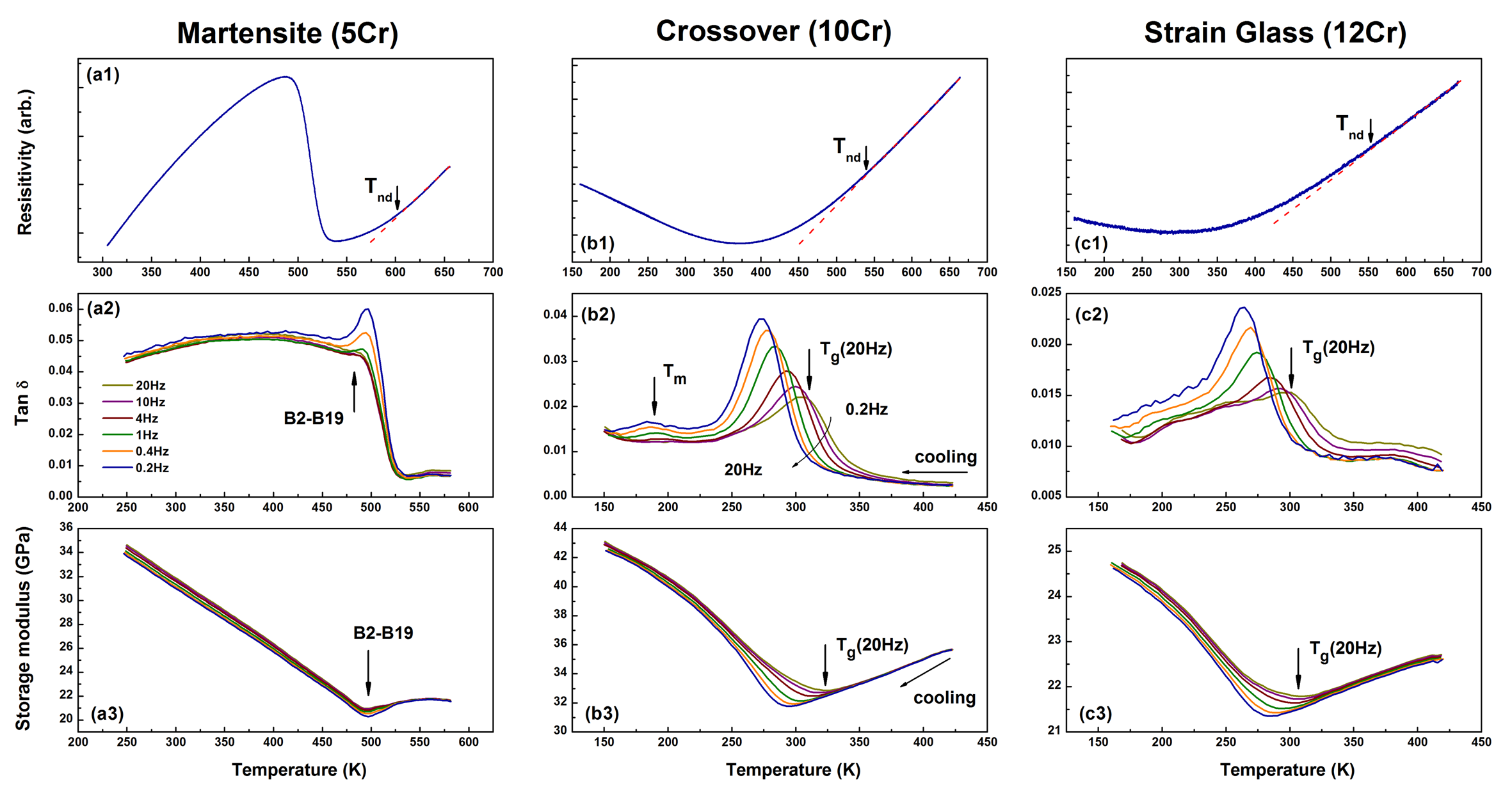}

\caption{The transformation behavior of low Cr-content Ti$_{50}$(Pd$_{45}$Cr$_{5}$) alloy (a1)(a2)(a3), 
intermediate Cr-content Ti$_{50}$(Pd$_{40}$Cr$_{10}$) alloy (b1)(b2)(b3) and high Cr-content Ti$_{50}$(Pd$_{38}$Cr$_{10}$)  alloy (c1)(c2)(c3), by means of electrical resistivity, DMA. } 
\label{Fig4}
\end{center}
\end{figure*}

\begin{figure}[tbph] 
\begin{center}
    \includegraphics[width=8.0cm]{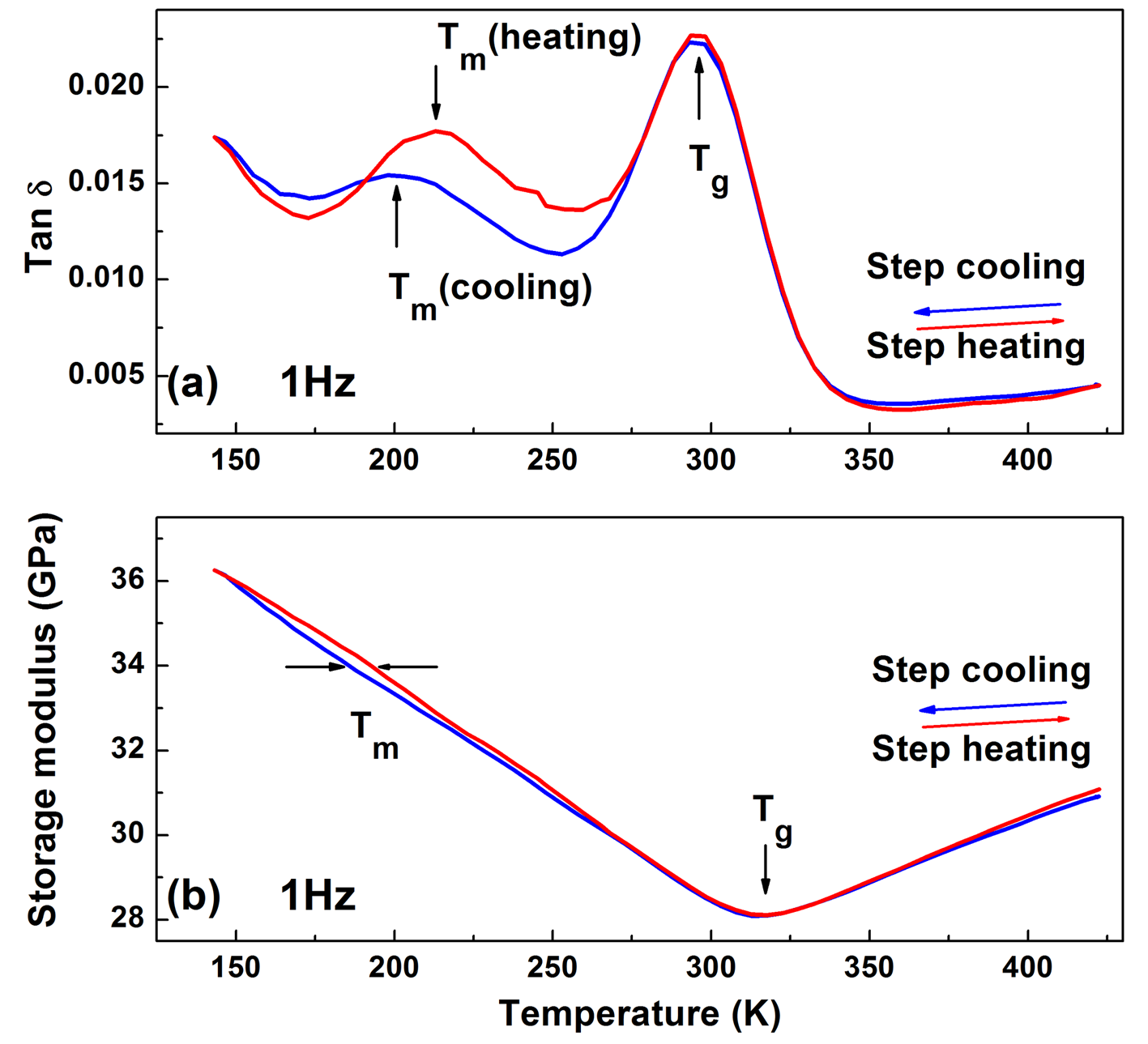}

\caption{DMA results on step cooling and step heating processes of Ti$_{50}$(Pd$_{40}$Cr$_{10}$) alloy, under frequency of $1Hz$. } 
\label{Fig5}
\end{center}
\end{figure}

\begin{figure}[tbph] 
\begin{center}
    \includegraphics[width=8.0cm]{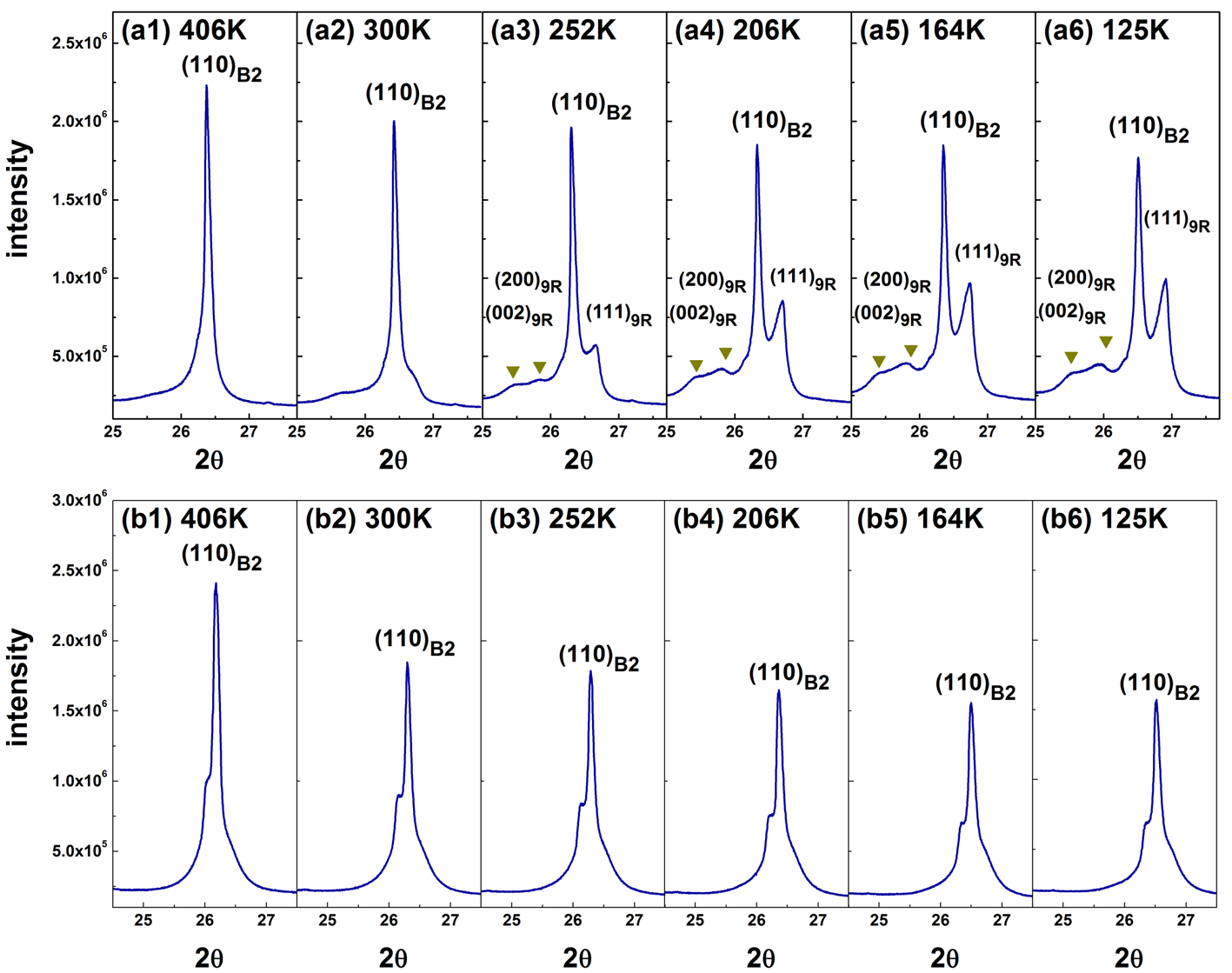}

\caption{In-situ sychrotron XRD patterns of (a) Ti$_{50}$(Pd$_{40}$Cr$_{10}$) alloy and (b) Ti$_{50}$(Pd$_{38}$Cr$_{12}$) alloy from $400K$ to $125K$. } 
\label{Fig6}
\end{center}
\end{figure}

\begin{figure}[tbph] 
\begin{center}
    \includegraphics[width=8.0cm]{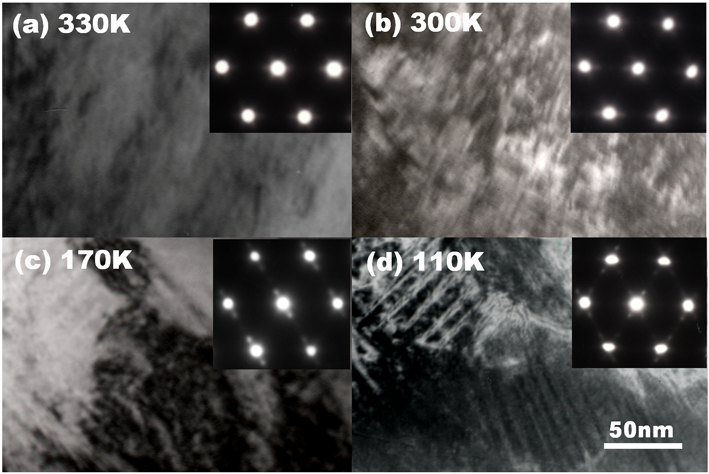}

\caption{In-situ TEM observation of Ti$_{50}$(Pd$_{40}$Cr$_{10}$) alloy from $330K$ to $110K$. } 
\label{Fig7}
\end{center}
\end{figure}

To check the predictions of our RG calculations, we experimentally investigated the phase transformation 
behavior as function of temperature $T$ and concentration $x$ for Ti$_{50}$(Pd$_{50-x}$Cr$_x$) alloys.

In the low Cr-content regime ($x \leq 8$), the system undergoes a B2$\rightarrow$B19 martensitic transformation. 
The transformation properties of Ti$_{50}$(Pd$_{45}$Cr$_5$) alloy are shown in Figs.~\ref{Fig4}(a1)-~\ref{Fig4}(a3). 
The B2$\rightarrow$B19 martensitic transformation at its transformation temperature $M_s$ is accompanied by a sharp 
increase in electrical resistivity [Fig.~\ref{Fig4}(a1)]. It is found that the electrical-resistivity curve deviates from linearity above $M_s$, and the onset temperature of the deviation $T_{nd}$ is defined as the tweed formation temperature. The martensitic transformation is also characterized by a frequency-independent peak in internal friction [Fig.~\ref{Fig4}(a2)] and a frequency-independent dip in storage modulus [Fig.~\ref{Fig4}(a3)]. Note that the frequency independence in the Dynamical Mechanical Analysis (DMA) dip/peak is an important feature of martensitic transformation. 

For high Cr-content ($x \geq 12$), the alloys adopt a transition path from austenite through tweed to strain glass. The transformation properties of Ti$_{50}$(Pd$_{38}$Cr$_{12}$) alloy are shown in Figs.~\ref{Fig4}(c1)-~\ref{Fig4}(c3). As shown in Fig.~\ref{Fig4}(c1), the electrical resistivity also shows a deviation from linearity below $T_{nd}$ , indicating the appearance of tweed state. The alloy undergoes a frequency dependent storage modulus dip and an internal friction peak [Figs.~\ref{Fig4}(c2) and (c3)], in contrast with the frequency independent behaviors during the martensitic transformation [Fig.~\ref{Fig4}(a1)]. This demonstrates a dynamic freezing transition strain glass transition occurs in high Cr-content alloys. The ideal glass frozen temperature ($T_0$)  was obtained from  by fitting the  DMA dip temperature $T_g(\omega)$  and frequency $\omega$ with the Vogel--Fulcher 
relation $\omega = \omega_0 \exp \left[-E_a /k_B (T_g(\omega)-T_0) \right] $. 
Note that the frequency dependence in the DMA dip/peak is an important feature of strain glass transformation.

For the crossover regime ($9 < x < 12$) between martensite and strain glass, the alloys experience all four strain states 
upon cooling and the spontaneous transformation from strain glass to martensite phase (9R) takes place. Figs.~\ref{Fig4}(b1)-4(b3) show the predicted spontaneous transformation behavior for Ti$_{50}$(Pd$_{40}$Cr$_{10}$) alloy. Upon cooling, the tweed first form at $T_{nd}=531K$ [Fig.~\ref{Fig4}(b1)]. Further cooling gives rise to a frequency dispersive internal friction peak and a storage modulus dip in the DMA results in Figs.~\ref{Fig4}(b2) and (b3), which correspond to a strain glass transition with frozen temperature $T_0$  of $250K$. With a further decrease in temperature, a frequency-independent internal friction peak appears in the DMA results [Fig.~\ref{Fig4}(b2)], which shows the similar feature as the martensitic transformation [Fig.~\ref{Fig4}(a2)]. This indicates a certain phase transformation occurs. We further studied the thermal hysteresis of the two phase transformations to characterize them. As shown in Fig.~\ref{Fig5}, the glass transition at higher temperature is associated with nearly zero thermal hysteresis in the internal friction anomaly and storage modulus, which is consistent with previous studies
on the strain glass transition~\cite{Ref11,Ref12,Ref13,Ref14,Ref15}. From the peak interval of internal friction and storage modulus in Fig.~\ref{Fig5}, we found the lower temperature transition are associated with a thermal hysteresis about $10K$, which shares the same feature with a first-order transition. We interpret the transition as a spontaneous transformation from strain glass to normal martensite transformation. We also employed in-situ synchrotron XRD to check the predicted spontaneous behavior for Ti$_{50}$(Pd$_{40}$Cr$_{10}$), as shown in Fig.~\ref{Fig6}(a). In the tweed region ($400K$ and $300K$), only a sharp $(110)_{B2}$ peak appears without peak splitting. At $252K$ where strain glass is frozen, clear peak splitting can be observed and the new peaks can be indexed as $(002)_{9R}$,$(200)_{9R}$ and $(111)_{9R}$, suggesting presence of short range order associated with the 9R structure. With further decrease to $125K$, the peak splitting becomes stronger and the 9R peak height drastically increases, indicating the formation of 9R martensite. To exclude the possibility of the above peak splitting is caused by nanodomain growth,  contrast in-situ synchrotron XRD was also carried our on Ti$_{50}$(Pd$_{38}$Cr$_{12}$), in which the strain glass state is stable, but the system can also show the nanodomain growth with the decrease of temperature~\cite{Ref11,Ref12,Ref13}. Fig.~\ref{Fig6}(b) clearly shows that the $(110)_{B2}$ peak keeps no splitting throughout the measurement temperature range. Therefore, the peak splitting in Ti$_{50}$(Pd$_{40}$Cr$_{10}$) alloy should indicate  a  transformation from strain glass to the martensite.

The microstructure evolution in Ti$_{50}$(Pd$_{40}$Cr$_{10}$)was further investigated by in-situ TEM observations from $330K$ to $110K$, shown in Fig.~\ref{Fig7}. In the tweed region($330K$ and $300K$), contrast from nano-domains can be seen with diffused super spots in the diffraction pattern; however, when cooling to $110K$ (martensite region), at which 9R peaks appears in the Bragg reflections, parallel martensite domains are visible in addition to nanodomains. From the present experimental results, we conclude that the second-step transition is the spontaneous transition from strain glass to normal martensite phase 9R. We note that although the spontaneous transition does occur, the parent peak still exists in the synchrotron XRD pattern. This suggests that not all the nano-domains in strain glass spontaneously transform into 9R martensite within our measurement window. The lack of completeness of such a spontaneous transition can be ascribed to kinetic limitations. In addition, it is also noted that the spontaneous phase transition here is rather weak and sluggish as the phase mixture of strain glass and martensite exists over a wide temperature range. This could be the reason for the absence of anomaly in our conventional DSC measurement (in which only latent heat is measured). Changes in heat capacity may aid in identifying the spontaneous transition. 

According to the above experimental results, a modification is made to the previous phase diagram of Ti$_{50}$(Pd$_{50-x}$Cr$_x$) alloys~\cite{Ref36}, where a crossover composition regime is included, as shown in Fig.~\ref{fig1}. Similar phase diagram including spontaneous phase transition, can be found in Ti$_{50}$(Ni$_{50-x}$Fe$_x$) strain glass alloy~\cite{Ref37}, and also in La-modified Pb(Zr$_{0.65}$Ti$_{0.35}$)O$_3$ ferroelectric relaxor ceramic~\cite{Ref38}.

\section{Conclusion}

\label{SectionGeneralization}

Our study emphasizes the importance of statistical mechanics and spin glass theory which, in conjunction with experiments, provide a general 
framework to understand universal features of the strain glass and tweed phases in ferroelastic materials. 
Our specific aim has been to emphasize how pseudo spin models
of martensites provide a predictive route towards understanding aspects of glass behavior seen in experiments.
This suggests that ferroelastics are very similar to ferroelectric and ferromagnetic materials, in the sense that
they can be described within the same framework of statistical mechanics of spin models. A crucial feature
of ferroelastic spin models is the additional $S=0$ state, which allows for two different ``paraelastic''
phases, austenite and tweed. Tweed is characterized in our study as a disordered phase of martensitic variants.
This tends to show that the tweed state observed is a true thermodynamical phase, that exists even in the
absence of long-range interactions, just like martensite and 
austenite. This is an important outcome of our study as so far tweed has been mostly interpreted as a spin glass or as 
a static precursor.
We also believe that many key ideas emanating from the spin glass/statistical mechanics community
should apply to the case of ferroelastics as well. For example, it still seems widely believed in the strain glass community 
that the precise form of the quenched disorder, modeling impurities, is crucial in understanding the physics of the systems under scrutiny.
We know from experience that this is not true for spin glasses, and it seems highly unlikely that it could be true for disordered
ferroelastics. We also wish to emphasize that long-range interactions, though important for microstructures, 
do not seem to be relevant to understand the global topology of the phase diagram.  

Although we have mainly focused on a specific model in 2D, we argued in Sec.~\ref{secGene3D}
that our conclusions should apply to a wide variety of 2D and 3D ferroelastic transitions;
in particular, we believe that our calculated phase diagram is ``generic''. 
To be more precise, we expect to find 
very similar topology of phase diagrams for other transitions with more variants in both two and three spatial dimensions. 
The alloy we chose as an example in the paper gives rise to two product phases as a function of disorder (B2 to B19, and B2 to 9R). 
Our analysis predicts that ferroelastics undergoing transitions to one product phase, such as TiNiFe, FePd or CaTiO$_3$, will show a very similar phase diagram and a spontaneous transition. We thus expect the general topology of the phase diagram shown in Fig.~\ref{FigMC}(3) to be quite robust.

\begin{acknowledgments} 
We thank Gao Jinghui and Zhang Zhen for help with the experiments. D.X. and X. R. are grateful to Spring-8 for the use of beamline BL15XU where the synchrotron XRD measurements were made. R.V. and W.E. wish to thank A. Lazarescu for useful discussions.
R.V. and X.D. are grateful to the Theoretical division and CNLS, LANL for support. This work was supported by the US DOE at LANL (DE-AC52-06NA25396) as well as the NSFC of China (Grant Nos. 51171140, 51231008), the 973 project of China under Grant No. 2010CB631003 and the 111 project of China.
\end{acknowledgments}

\end{document}